\documentclass[pdftex,epjc3,twocolumn]{svjour3} 

\RequirePackage{fix-cm} 
\smartqed  
\RequirePackage{graphicx}
\usepackage{graphicx}
\usepackage{bm}
\usepackage{soul}
\usepackage{amssymb}
\usepackage{amsmath}
\usepackage{amsfonts}
\usepackage{xcolor}
\newcommand \sqn{$\sqrt{s_{_{NN}}}$ } 
\journalname{Eur. Phys. J. C}
\begin{document}


\title{Scaling properties of direct photon yields \\
in heavy ion collisions }

\author{Vladimir Khachatryan\thanksref{e1,addr1} \and  \\ Micha{\l} Prasza{\l}owicz \thanksref{e2,addr2}}
\thankstext{e1}{e-mail:vladimir.khachatryan@stonybrook.edu}
\thankstext{e2}{michal@if.uj.edu.pl}

\institute{Department of Physics and Astronomy, Stony Brook University,  Stony Brook, New York 11794-3800, USA \label{addr1}
\hspace{-0.8cm} \and 
Institute of Theoretical Physics, Jagiellonian University,  S. {\L}ojasiewicza 11,
30-348 Krak{\'o}w, Poland\label{addr2}
}


\date{Received: date / Accepted: date}
\maketitle

\begin{abstract}
A recent analysis from the PHENIX collaboration of available direct photon measurement results  
in collisions of various systems such as Au+Au, Cu+Cu, and Pb+Pb,
at different beam energies ranging from 39 to 2760~GeV,
has shown a universal, within experimental uncertainties,  {\em multiplicity} 
scaling, in which  direct photon $p_{T}$-spectra for transverse momenta 
up to 2 GeV/$c$  are scaled with charged hadron pseudorapidity 
density at midrapidity raised to power $\alpha=1.25$. On the other hand, 
those direct photon $p_{T}$-spectra also exhibit {\em geometrical} scaling in the similar $p_{T}$ range. 
Assuming power-law dependence of the scaled photon spectra for both scaling laws,
we formulate two independent conditions for  the power $\alpha$, which overshoot experimental 
data by $\sim 10\%$ on average. We discuss possible sources that might improve this estimate.
\end{abstract}


\section{\label{sec:intro}Introduction}

Measurements of direct photons provide unique opportunities in probing and studying the properties 
and evolution of the matter produced in heavy ion collisions (HIC). These photons are defined to be produced from all 
the sources except for hadronic decays. Since they hardly interact with the ``fireball'' of quarks and gluons due to a small 
interaction cross section with the medium, the information they carry from the time of their production is not washed out 
by final state interactions. Experimentally measured low momentum direct photon $p_{T}$-spectra by PHENIX 
(in Au+Au at \sqn = 200~GeV and 62.4~GeV, in Cu+Cu at \sqn = 200~GeV)
\cite{Adare:2008ab,Afanasiev:2012dg,Adare:2014fwh,Adare:2018jsz,Adare:2018wgc,Khachatryan:2018evz,Khachatryan:2018uzc}
and ALICE (in Pb+Pb at \sqn = 2760~GeV) \cite{Adam:2015lda}
collaborations in HIC are enhanced with respect to $N_{\rm coll}$ (number of binary nucleon collisions) scaled 
reference yield (measured or calculated) in p+p collisions.
Low momentum direct photon measurements by STAR collaboration show less enhancement
 \cite{STAR:2016use}. Earlier low energy WA98 data \cite{Aggarwal:2000th} have mostly upper bounds in the 
relevant $p_T \le 2$~GeV$/c$ region. Direct photons in HIC also show large anisotropy (elliptic flow) 
\cite{Khachatryan:2018uzc,Adare:2015lcd,Acharya:2018bdy}.

\begin{figure}[h]
\centering
\includegraphics[scale=0.435]{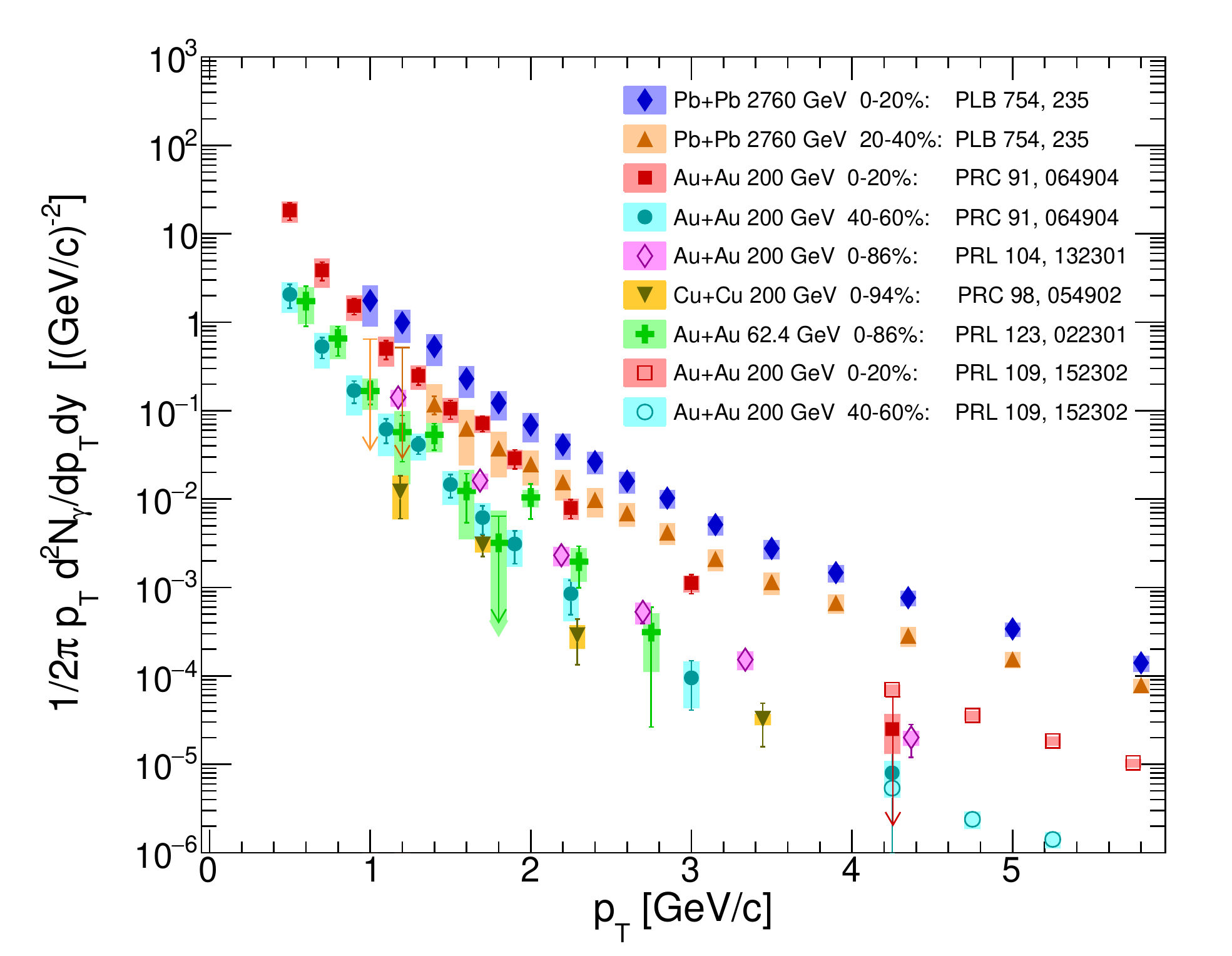}
\caption{Direct photon invariant yield as a function of $p_{T}$ for various colliding nuclei and collision 
centrality selections at three center-of-mass energies. The plot labels give the references of the shown data. 
The vertical lines of the data points describe statistical uncertainties, the boxes describe systematic uncertainties.
}%
\label{fig:unscaled}%
\end{figure}

Fig.~\ref{fig:unscaled} shows several data sets of direct photon $p_{T}$-spectra at low- ($<$ 1 GeV$/c$) and 
intermediate- (from 1~GeV$/c$ up to $\sim $ 5~GeV$/c$) $p_{T}$ regions. There have been many theoretical 
attempts to reproduce the photon yields shown in Fig.~\ref{fig:unscaled} and flow coefficients (that are not 
discussed here) with, however, mixed success. Hydrodynamical simulations of the fireball evolution 
\cite{DPSYJG:2011,Shen:2013cca,SHPG:2014,Paquet:2015lta}, calculations in the framework of the elliptic-fireball
expansion scenario \cite{vanHees:2011,Rapp:2014,vanHees:2015}, Parton-Hadron-String Dynamics transport 
approach \cite{Bratkovskaya:2008,Bratkovskaya:2014,Linnyk:2014,Linnyk:2015}, as well as the spectral function 
approach \cite{Dusling:2010,Duslingthesis,Lee:2014,Kim:2017},
have encountered difficulties in the simultaneous description of the observed large yields and 
large anisotropies, which have given rise to a challenging problem, commonly referred to as {\em direct (thermal) 
photon flow puzzle} \cite{Shen:2016odt} (for recent review see \cite{David:2019wpt}).
 
The early attempts to describe the photon yields by initial state models can be found in 
Refs.~\cite{Chiu:2012ij,Klein-Bosing:2014uaa,McLerran:2014oea,McLerran:2014hza,McLerran:2015mda}. 
More recently initial state models have been used both as a part of hydrodynamic evolution \cite{Paquet:2015lta} 
and in a bottom-up thermalization scenario \cite{Berges:2017eom,Khachatryan:2018ori}. In the latter case
\cite{Khachatryan:2018ori} good fits for the photon yields have been obtained for PHENIX and ALICE data 
(the anisotropies are not discussed there).

Much work still needs to be done to include the anisotropies in the initial state model calculations, in order 
to directly address the aforementioned puzzle. Ref.~\cite{McLerran:2014hza} discusses the possibility of late stage 
elliptic flow effects to be explained in the combined Glasma and thermal photon production scenarios, by which the 
photon mean emission time is shifted toward a later time scale as compared to the mean emission time obtained only 
from the thermal ansatz. Nonetheless, as it has been pointed out in \cite{Khachatryan:2018ori}, a more realistic 
description of the photon production might come out within the combined framework of the initial and late stage
models. In this regard a promising step is already undertaken in \cite{Paquet:2015lta}, where the event-by-event 
hydrodynamical model uses IP-Glasma initial state conditions.

\begin{figure}[h]
\centering
\includegraphics[scale=0.435]{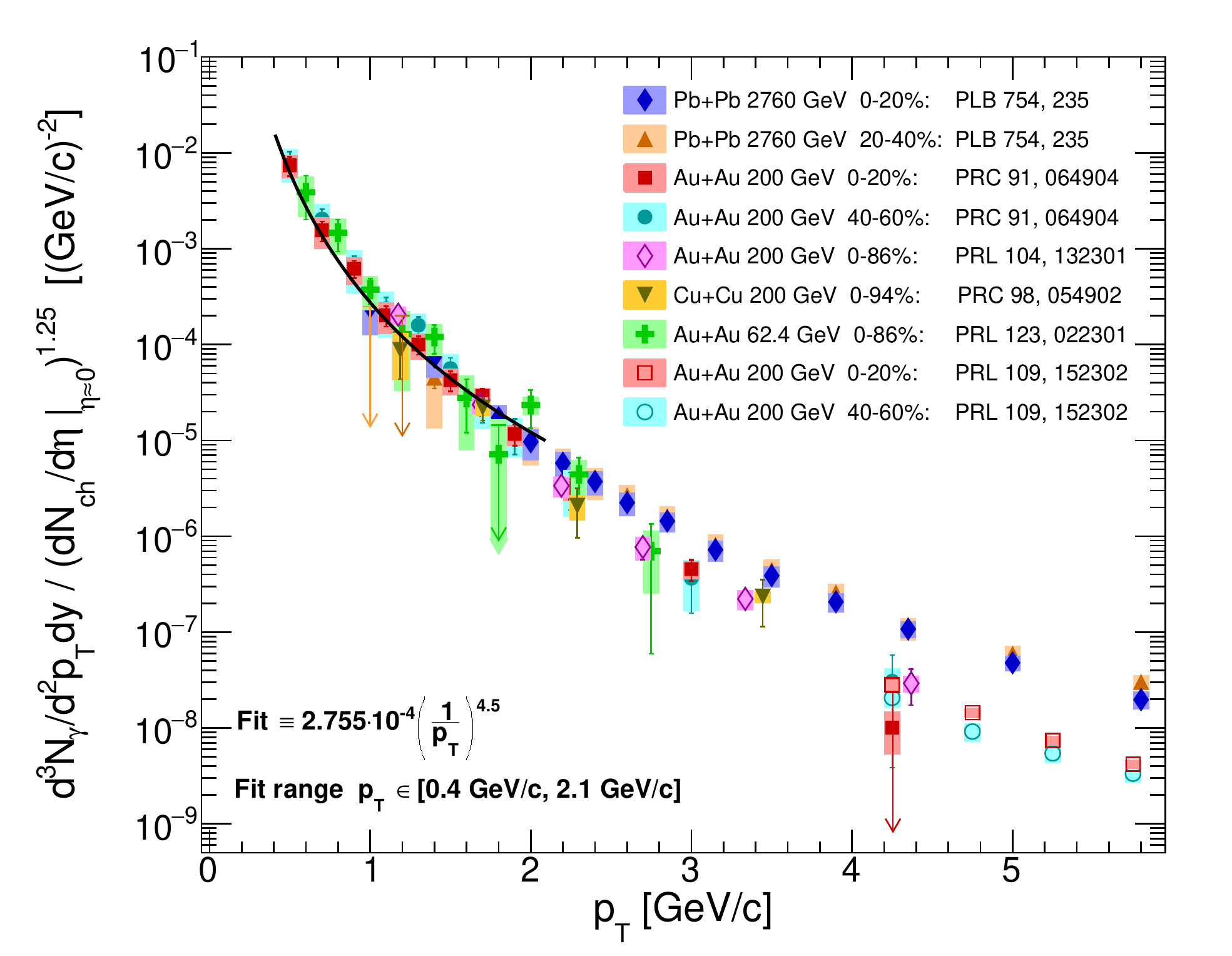}
\caption{Direct photon $p_{T}$-spectra from Fig.~\ref{fig:unscaled} scaled by ${(dN_{\rm ch}/d\eta)}^{1.25}$. 
The black curve corresponds to a power-law fit, the details of which are shown in the plot. See the next sections for 
more details.
}%
\label{fig:gammaMS}%
\end{figure}

However, the most important argument from \cite{Chiu:2012ij,Klein-Bosing:2014uaa,Khachatryan:2018ori}, directly 
applicable to the studies of this paper, indicates that the initial state models potentially allow the photon $p_{T}$-spectra 
to exhibit geometrical scaling (GS)\footnote{We will perform detailed studies of GS of photon $p_{T}$-spectra in HIC 
elsewhere \cite{KhaPrasz}.}. In fact, two types of scalings  are observed in direct photon 
$p_{T}$  distributions in HIC and hadron collisions:

\paragraph{Direct photon multiplicity scaling (MS).} It is an experimental observation in HIC, which 
shows that direct photon invariant yields follow one universal curve within experimental statistical and systematic uncertainties 
for various colliding species and system sizes at different center-of-mass energies, when scaled by charged hadron multiplicities 
at midrapidity raised to  the power of $\alpha=1.25$ \cite{Adare:2018wgc}:
\begin{equation}
\frac{1}{\left( dN_{\mathrm{ch}}/d\eta|_{\eta \approx 0}\right )^{\alpha}}
\frac{dN_{\gamma}}{d^2p_{T}dy} = \frac{1}{Q_{0}^{2}}G(p_{T}),
\label{MS}%
\end{equation}
where $G$ is a universal energy- and multiplicity-inde{\-}pen{\-}dent function of $p_{T}$ 
(see Fig.~\ref{fig:gammaMS})\footnote{We will henceforth simply use $dN_{\mathrm{ch}}/d\eta$, 
skipping the midrapidity notation $\eta \approx 0$ in the text, and keeping it only in the figures.}, and $Q_{0}
\sim 1$~GeV/$c$. 
MS holds for small and intermediate $p_T$  up to $\sim 2 \div 2.5$~GeV$/c$,
precisely in the enhancement region discussed above.

\paragraph{Geometrical scaling (GS).} Here the invariant yields of direct photons (and also of charged hadrons)
can be related to a universal energy-independent function of scaling variable $\tau$ 
\cite{Klein-Bosing:2014uaa,KhaPrasz,McLerran:2010ex,McLerran:2011,Praszalowicz:2011rm,Praszalowicz:2018vfw}:
\begin{equation}
\frac{1}{S_{T}}\frac{dN_{\gamma,\text{ch}}}{d^2p_{T}d \eta}%
=\,F_{\gamma,\text{ch}}(\tau),
\label{GS}%
\end{equation}
where $S_{T}$ is a parameter characterizing geometrical overlap area of colliding nuclei. The variable $\tau$ is given by
\begin{equation}
\tau = p_{T}/Q_{\mathrm{s}}(x) \label{tau}.%
\end{equation}
GS has been for the first time observed in deep inelastic scattering (DIS) in e+p collisions 
\cite{Stasto:2000er}. The quantity  $Q_{\mathrm{s}}(x)$ in (\ref{tau}) is the saturation scale, {\em i.e.}, 
the transverse momentum at which  the number of gluons 
with longitudinal momentum fraction $x$, called Bjorken variable, cannot grow any more due to the nonlinearity of QCD 
evolution. This happens at gluon occupancy numbers being of the order of $1/\alpha_{s}$ ($\alpha_{s}$ -- strong coupling 
constant).
The phenomenon of gluon saturation is generated from such dynamics 
\cite{Gribov:1984tu,Mueller:1985wy,BK,Kovchegov1,Kovchegov2,Mueller:2002zm,Munier:2003vc,MPesch}
 and can be described by the Color Glass Condensate (CGC) effective theory of high energy scattering
\cite{McLerran:1993ni,McLRV2,McLRV2,elis:2010nm,McLerran:2010ub}. 

One still should remember that GS is more general. Indeed, it has been shown in Ref.~\cite{Caola:2008xr}
that  the approximate GS is in fact a general property of solutions of the QCD collinear evolution with general boundary 
conditions.  Furthermore, the results of Ref.~\cite{Iancu:2002tr} indicate that the rapidity evolution preserves 
GS beyond saturation when starting from the boundary conditions in the saturation region. It means that the scaling 
properties of an initial state are preserved and even built up by the QCD evolution. This may explain the fact that 
the geometrical scaling is observed in the hadronic $p_T$-spectra and even more so in the case of photons.

\begin{figure}[h]
\centering
\includegraphics[scale=0.435]{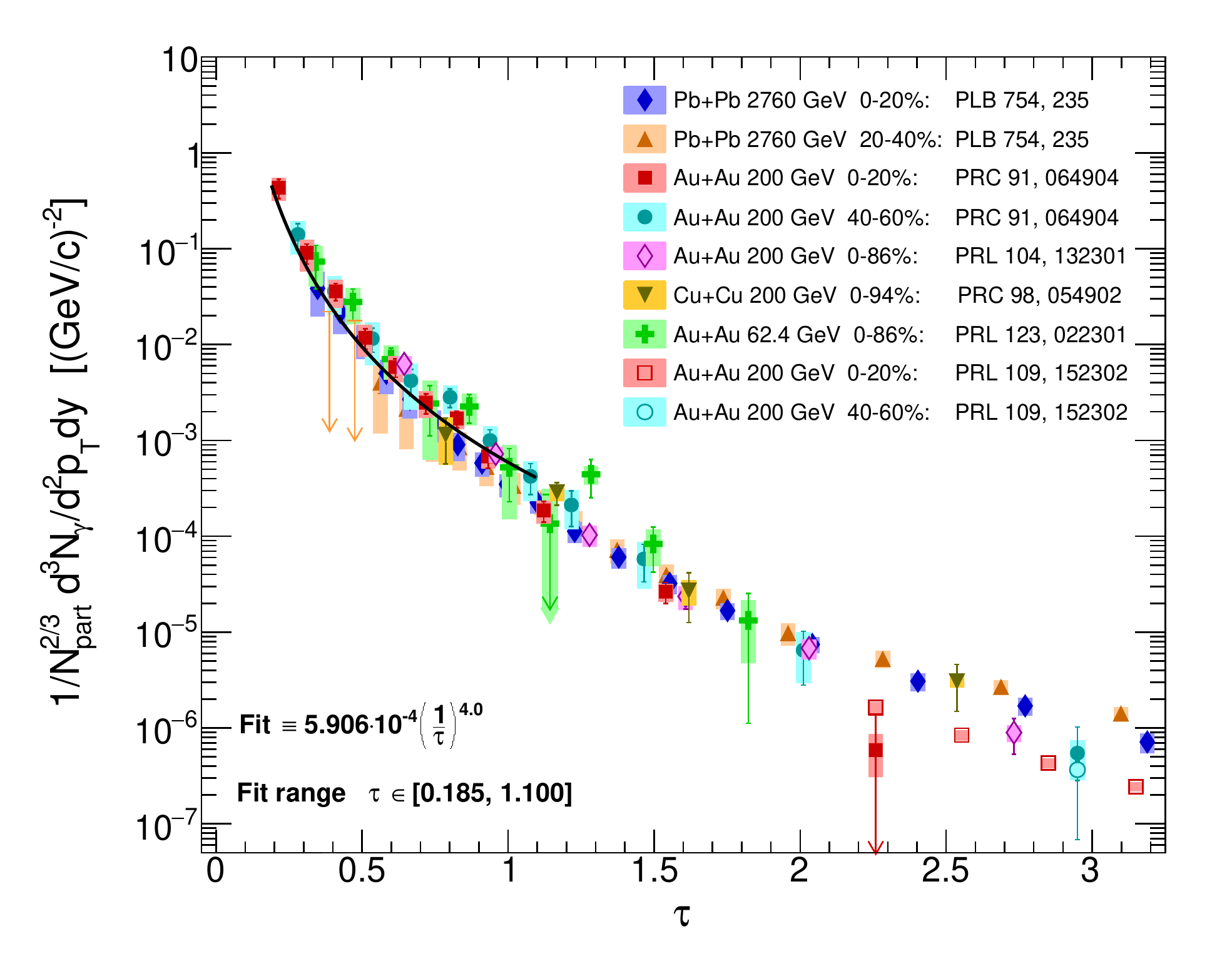}
\caption{Direct photon $p_{T}$-spectra from Fig.~\ref{fig:unscaled} exhibit geometrical scaling
when plotted in terms of  the scaling variable $\tau$ (at $\lambda \approx 0.2$) and divided by $S_{T} \sim N_{\rm part}^{2/3}$.
The black curve corresponds to another power-law fit, the details of which are shown in the plot. See the next sections 
for more details.
}%
\label{fig:gammaGS}%
\end{figure}

Saturation models predict specific dependence of $Q_{\mathrm{s}}$
on the collision energy, transverse momentum and collision centrality class, quantified by
the number of participating nucleons
\begin{equation}
Q_{\rm{s}} = N_{\rm{part}}^{\delta/2}\,Q_{0}\left( \frac{p_{T}}{ \sqrt{s_{_{\!NN}}}\times 10^{-3}}\right)^{-\lambda/2},
\label{Qsat1}
\end{equation}
with $\delta \sim 2/3$ and $\lambda \sim 0.2 \div 0.35$ \cite{Praszalowicz:2018vfw}.
Here we have chosen $x_0=10^{-3}$ for a typical value of Bjorken $x$ where the saturation effects
become important.
The direct photon GS is illustrated in Fig.~\ref{fig:gammaGS}.

In the present paper we formulate two conditions that allow  one to calculate the power $\alpha$ of MS in 
terms of the parameters that enter  the theoretical parametrization of GS, assuming power-law dependence 
of the scaled photon spectra. These two conditions result in two predictions for $\alpha$: 1.34 and 1.40, 
{\em i.e.} $\sim$ 10\% above the experimental value. Given the simplicity of the present analysis and approximate nature 
of both scaling laws, this result may be considered as a satisfactory one. Nevertheless, we also discuss possible corrections 
that might improve this theoretical prediction of $\alpha$.

\section{\label{sec:basics}Basics of Geometrical Scaling}

 Geometrical Scaling is based upon the assumption
that bulk properties of charged particles and photons share scaling properties encoded in the initial state wave 
function of  colliding hadrons/nuclei. While detailed models of direct photon production 
in the early and intermediate stages of HIC can be found in 
Refs.~\cite{Chiu:2012ij,McLerran:2014oea,McLerran:2014hza,Berges:2017eom,Khachatryan:2018ori}
as it was noted in the introduction, here we discuss general assumptions and necessary conditions leading to GS.

Differential gluon production cross-section 
can be written in terms of the unintegrated gluon distributions \cite{Gribov:1981kg}:
\begin{equation}
\frac{d\sigma}{dyd^{2}p_{\mathrm{T}}} = \frac{C}{p_{{T}}^{2}}%
{\displaystyle\int}
d^{2}\vec{k}_{T}\,\alpha_{\mathrm{{s}}}
\varphi^{(1)}(x_{1}%
,\vec{k}_{T}^{2})\,\varphi^{(2)}(x_{2},(\vec{k}-\vec{p})_{T}^{2}),
\label{Nchdef}%
\end{equation}
where $C$ contains color factors and numerical constants. The Bjorken $x$'s of colliding partons read
\begin{equation}
x_{1,2} = e^{\pm y}\,{p_{T}}/{\sqrt{s}}\,.
\end{equation}
%
In the midrapidity region $y\approx 0$, hence $x_1=x_2=x$.

There exist many models of unintegrated gluon distributions $\varphi(x,\vec{k}_{T}^{2})$ (see \emph{e.g.}
\cite{GolecBiernat:1998js,GolecBiernat:1999,Kharzeev:2002ei,Szczurek:2003vq}) that enter (\ref{Nchdef}). Most of them share 
two important features: geometrical scaling and dependence on the transverse area parameter $A_{T}$:
\begin{equation}
\varphi(x,\vec{k}_{\mathrm{{T}}}^{2}) = A_{T}\,\phi(k_{T}^{2}%
/Q_{\mathrm{{s}}}^{2}(x)),
\label{unintg}
\end{equation}
where $\phi$ is a dimensionless function of the scaling variable $\tau^{2} = %
k_{T}^{2}/Q_{\mathrm{{s}}}^{2}(x)$. The precise meaning of $A_{T}$ is best
understood in a picture where the impact parameter is also taken into 
account~\cite{Kowalski:2003hm,Tribedy:2010ab}.

Ignoring momentum dependence of the strong coupling constant, on dimensional analysis grounds, 
we arrive at
\begin{equation}
\frac{d\sigma}{dyd^{2}p_{{T}}} = A^{(1)}_{T}A^{(2)}_{T}{F}(\tau),
\label{sigmaGS}%
\end{equation}
where ${F}$ is a universal, energy-independent function of the scaling variable $\tau$ 
in (\ref{tau}).

If one assumes that $A_{T}$ is an energy-independent constant, which is true in the case of 
the GBW model \cite{GolecBiernat:1998js,GolecBiernat:1999} for DIS, then it is the differential cross-section that 
should exhibit GS. Indeed, it has been shown in Ref.~\cite{Praszalowicz:2015dta} that for charged 
particles the differential cross-section  in p+p scattering 
scales better (over larger $p_{T}$ interval) than multiplicity, and the exponent 
$\lambda \approx 0.3$ is compatible with the DIS scaling \cite{Praszalowicz:2012zh}.
In contrast, for multiplicity scaling $\lambda \approx 0.2$.

 In order to obtain the multiplicity distribution, we should divide both sides of 
(\ref{sigmaGS}) by an appropriate inelastic cross-section:
\begin{equation}
\frac{dN}{dyd^{2}p_{{T}}} = \frac{A^{(1)}_{T}A^{(2)}_{T}}{\sigma_{\mathrm{inel}}}%
{F}(\tau). \label{sigma_vs_N}%
\end{equation}
This distribution would scale if
\begin{equation}
S_{T}={A^{(1)}_{T}A^{(2)}_{T}}/{\sigma_{\mathrm{inel}}}%
\label{ST1}
\end{equation}
were energy-independent. This assumption leads to (\ref{GS}). 
It is the energy dependence of $S_{T}$ that
is responsible for different scaling properties of the cross-section and multiplicity  of charged hadrons
in p+p collisions  \cite{Praszalowicz:2012zh}.
In the case of  HIC for a fixed centrality, $S_{T}$ has geometrical interpretation as an 
overlap area of two colliding nuclei~\cite{Kharzeev:2002ei,Kharzeev:2000ph}
that scales like $N_{\mathrm{part}}^{2/3}$.
Throughout this paper we shall assume $Q_{\rm s}\sim S_{T}\sim N_{\mathrm{part}}^{\delta}$. 
Possible weak energy dependence of $S_{T}$ would lead to the violation of geometrical scaling.
 For small systems (like  $e.g.$ dA or pA) one has two saturation scales that differently
scale with $N_{\rm part}$ and the  effective saturation scale has to be taken as 
$\sqrt{Q_{\rm{sat}}[{\rm large}]Q_{\rm{sat}}\left[{ \rm{small}}\right] }$ \cite{Klein-Bosing:2014uaa,Dumitru:2001ux}. 
We address this issue in \cite{KhaPrasz}. Here we concentrate on large systems only.

The functional form of the saturation scale $Q_{\mathrm{{s}}}(x)$ in (\ref{Qsat1}) follows from the 
nonlinear QCD evolution \cite{Munier:2003vc}, and in the case of heavy ions, also 
from the collision geometry \cite{Kharzeev:2002ei,Kharzeev:2000ph}, resulting in the following form
of the scaling variable $\tau$:
\begin{equation}
\tau = \frac{1}{N_{\text{part}}^{\delta/4}}\frac{p_{T}}{Q_{0}}\left(
\frac{p_{T}}{W}\right)^{\lambda/2},
\label{taugen}%
\end{equation}
%
where $W=\sqrt{s_{_{\!NN}}}\times 10^{-3}$ is an effective energy scale shown in (\ref{Qsat1}).

Should there be any 
energy dependence of $S_{T}$ (which enters the definition of $Q_{\rm s}$), it is already included in $\lambda$, 
as we fix $\lambda$ from the fits to data that do not distinguish different sources (QCD nonlinear evolution, 
$S_{T}$ energy dependence) of the energy increase of $Q_{\mathrm{s}}(W)$. These fits  
\cite{KhaPrasz}, 
as well as a direct inspection of Fig.~\ref{fig:gammaGS}, indicate that for direct photons $\lambda \approx 0.2$, 
rather than 0.3. The same conclusion has been reached in an earlier study of Ref.~\cite{Klein-Bosing:2014uaa}.

\section{\label{sec:Nch}Charged particle multiplicity}

\begin{figure*}[h]
\centering
\includegraphics[scale=0.745]{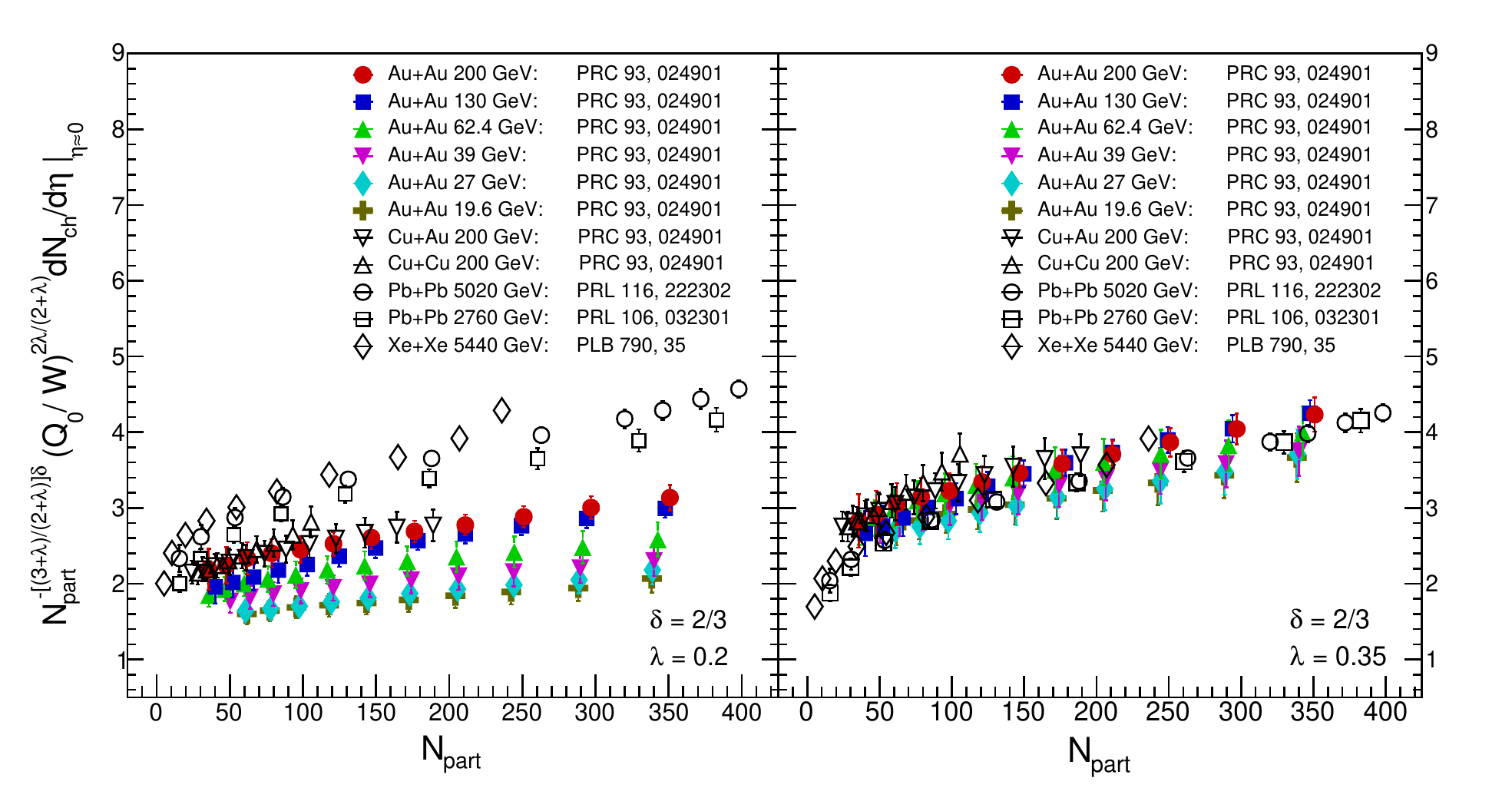}
\caption{Scaled multiplicity density (\ref{eq:scaledNch}) for charged particles at $\lambda=0.2$ and $\lambda=0.35$
as a function of $N_{\rm part}$.}%
\label{fig:Nch_scaling}%
\end{figure*}

Let us first calculate the charged particle multiplicity for large  collision systems from the scaling formula 
(\ref{GS}). To this end we will need a Jacobian to change  the integration variable from $p_{T}$
to $\tau$:%
\begin{equation}
p_{T}dp_{T}=\frac{2Q_{0}^{2}}{2+\lambda}N_{\text{part}}^{\frac{\delta}{2+\lambda}}
\left( \frac{W}{Q_{0}}\right )^{\frac{2\lambda}{2+\lambda}}%
\tau^{\frac{2-\lambda}{2+\lambda}}d\tau,
\end{equation}
which yields%
\begin{align}
\frac{dN_{\text{ch}}}{d \eta}  &  =%
{\displaystyle\int}
d^{2}p_{T}
\frac{dN_{\text{ch}}}{d^2p_{T}d\eta}
\nonumber\\
& = S_{T}\,Q_{0}^{2}\,\frac{2}{2+\lambda}N_{\text{part}}^{\frac{\delta
}{2+\lambda}}\left( \frac{W}{Q_{0}} \right)^{\frac{2\lambda}{2+\lambda}}%
{\displaystyle\int}
F_{\text{ch}}(\tau)\,\tau^{\frac{2-\lambda}{2+\lambda}}d\tau\nonumber\\
& = N_{\text{part}}^{\frac{3+\lambda}{2+\lambda}\delta}\left( \frac{W}{Q_{0}%
} \right)^{\frac{2\lambda}{2+\lambda}}\kappa,
\label{Nch}
\end{align}
where  the constant $\kappa$ includes the integral of $F_{\text{ch}}$,
$S_T$ and other irrelevant constants. 
Note that the energy dependence $\sim W^{2\lambda/(2+\lambda)}$ is compatible with 
multiplicity growth in p+p as measured at the LHC 
\cite{Aamodt:2009aa}. 
For small $\lambda$ and $\delta=2/3$ we have the quantity  
$N_{\text{part}}^{\frac{3+\lambda}{2+\lambda}\delta}\approx N_{\text{part}}$, 
as has been assumed {\em e.g.} in Ref~ \cite{Kharzeev:2000ph},
though it has been now established that better scaling of charged multiplicities
is obtained when $N_{\rm part}$ is replaced by the number of participating quarks
(see below).

In order to assess the quality of the scaling formula (\ref{Nch}), we plot the scaled charged
hadron multiplicity in Fig.~\ref{fig:Nch_scaling}: 
\begin{equation}
 N_{\text{part}}^{-\frac{3+\lambda}{2+\lambda}\delta}\left( \frac{Q_0}{W} \right)^{\frac{2\lambda}{2+\lambda}}
\, \frac{dN_{\text{ch}}}{d \eta}
\label{eq:scaledNch}
\end{equation}
for the data from Refs.~\cite{Adare:2015bua,Aamodt:2010cz,Adam:2015ptt,Acharya:2018hhy}
as a function of $N_{\rm part}$ for $\delta=2/3$ and for two different choices
of $\lambda$. We see that, in contrast to the photon  GS-scaled yields, the best alignment 
of all data is achieved for $\lambda\sim 0.35$. 
Here $Q_{0} \sim 1$~GeV/$c$ as before. Some systematic shift is seen in the case of
Cu+Cu data. We would expect that  the scaled multiplicity in (\ref{eq:scaledNch}) 
should not depend on $N_{\rm part}$, while we see that it rises
with $N_{\rm part}$. This is a well-known problem in HIC, and it has been argued that better scaling 
is obtained with the number of participating quarks \cite{Adare:2015bua}. 
Here it is enough to slightly change the value of $\delta$ from 2/3 to 3/4
 to ensure $N_{\rm part}$ independence of (\ref{eq:scaledNch}). However, 
as we shall see in the following, the value of $\delta$ does not enter the conditions for the MS exponent
 $\alpha$.

The fact that $\lambda$ for charged particle spectra is different than the one for photons is --
to our best knowledge -- not understood theoretically. It may be due to the interactions of charged 
particles in medium. The other reason might be that for charged particles we extract $\lambda$
from {\em integrated yields} that include the scaling part up to $2 \div 3$~GeV/$c$ \cite{Praszalowicz:2018vfw} and the 
non-scaling tail,  while in obtaining Eq.~(\ref{Nch}) we have assumed that GS
is present in the whole experimentally measured range. In the following we will accept this difference as a phenomenological observation,
and distinguish $\lambda$ for photons from $\lambda_{\rm ch}$ for charged particles.

\section{\label{sec:scaling}Relating the scaling laws}
Let us rewrite the scaling laws (\ref{GS}) and (\ref{MS}) in the following way
\begin{equation}
S_{T}\,F_{\gamma}(\tau(p_{T})) = \frac{dN_{\gamma}}{d^2p_{T}dy}
= \left( \frac{dN_{\text{ch}}}{d\eta} \right)^{\alpha} \frac{1}{Q_{0}^{2}}G(p_{T}),
\label{both}%
\end{equation}
which can be satisfied only when the left hand side and the right hand side
of Eq.\thinspace(\ref{both}) have the same $p_{T}$, $W$ and $N_{\text{part}}$ dependence.

It is clear that one cannot proceed further without explicit knowledge of  the functions $F_{\gamma}$ and 
$G$. To this end we shall assume a power-law dependence%
\begin{equation}
F_{\gamma}(\tau)\sim\frac{1}{\tau^{n}}~~~\mathrm{and}~~~G(p_{T})\sim\left(
\frac{Q_{0}}{p_{T}}\right)  ^{m}.
\label{eq:powerlaw}
\end{equation}
Indeed, as shown in Figs.~\ref{fig:gammaMS} and \ref{fig:gammaGS},  where we plot $F_{\gamma}$ and $G$ for 
Au+Au \cite{Adare:2008ab,Afanasiev:2012dg,Adare:2014fwh}, Cu+Cu \cite{Adare:2018jsz} and Pb+Pb
\cite{Adam:2015lda} direct photon data at different energies and centralities for $\lambda=0.2$ and $\delta=2/3$,
the power-law fall-off~\cite{Klein-Bosing:2014uaa} with $n,\, m\sim4,\, 4.5$ works pretty well at small and 
intermediate $\tau$ (or $p_T$). Note that $m=4$ is a generic prediction for radiation from the CGC 
\cite{Dumitru:2001ux} at intermediate $p_{T}$, while -- as we have checked by 
looking at prompt photons up to transverse momenta $\sim 2$~GeV$/c$ -- fits to the data prefer $m=5$.
Altogether
\begin{eqnarray}
& & \!\!\!\frac{dN_{\gamma}}{d^2p_{T}dy} \sim
N_{\text{part}}^{\delta}\left( N_{\text{part}}^{\delta/4}
\left( \frac{W}{Q_{0}} \right)^{\frac{\lambda}{2}}
\left( \frac{Q_{0}}{p_{T}} \right)^{\frac{2+\lambda}{2}}
\right)^{n}
\nonumber \\
& & ~~~\sim\left( N_{\text{part}}^{\delta} N_{\text{part}}^{\frac
{1}{2+\lambda_{\rm ch}}\delta}\left( \frac{W}{Q_{0}} \right )^{\frac{2\lambda_{\rm ch}
}{2+\lambda_{\rm ch}}}\right)^{\alpha}\left( \frac{Q_{0}}{p_{T}} \right)^{m},
\label{thesame}%
\end{eqnarray}
where we have used $S_{T} \sim N_{\mathrm{part}}^{\delta}$.  The functions in 
(\ref{thesame}) are proportional to each other if
\begin{equation}
m = \frac{2+\lambda}{2}\,n,~~\frac{4+n}{4}\,\delta=\frac{3+\lambda_{\rm ch}}{2+\lambda_{\rm ch}}%
\,\alpha \delta,~~\frac{\lambda}{2}\,n=\frac{2\lambda_{\rm ch}}{2+\lambda_{\rm ch}}\,\alpha.
\label{conditions}%
\end{equation}
From the first equation we see that the power-like fall-off should be faster for MS than for GS, and this 
prediction works quite well, as can be seen from Figs.~\ref{fig:gammaMS} and \ref{fig:gammaGS}.

Thereby, from (\ref{conditions}) we obtain the following:%
\begin{align}
\alpha  = \left. \frac{4+n}{4}\,\frac{2+\lambda_{\rm ch}}{3+\lambda_{\rm ch}}\right\vert
_{\substack{n=4\\ \lambda_{\rm ch}=0.35}}&=1.40, \label{first}\\
\alpha  = \left. n\,\frac{\lambda}{\lambda_{\rm ch}}\frac{2+\lambda_{\rm ch}}{4}\right\vert_{\substack{n=4\\ \lambda
=0.2\\ \lambda_{\rm ch}=0.35}} & = 1.34.
\label{second}%
\end{align}
Both estimates give  the power $\alpha$ compatible with the experimental value of 1.25 
(given our crude assumptions and approximations). Note that the first estimate follows from
the assumptions concerning the geometry of A+A collisions (although it 
does not depend on the actual value of $\delta$), whereas the second one follows from the energy 
dependence.

\section{Discussion and conclusions}

\begin{figure}[h]
\centering
\includegraphics[scale=0.38]{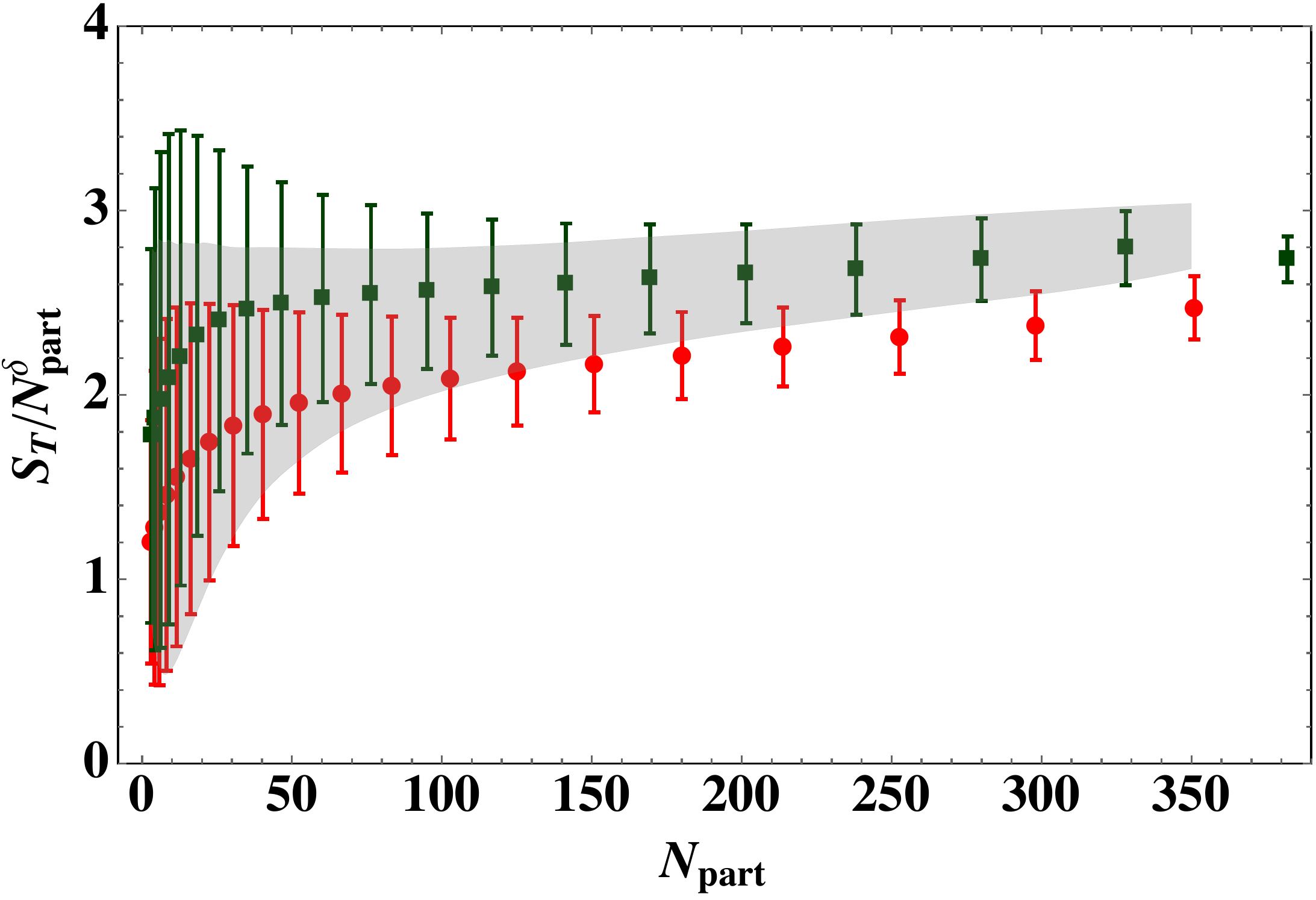}
\caption{
Glauber model predictions~\cite{Loizides:2017ack} for $S_{T}$ scaled by $N_{\mathrm{part}}^{\delta}$, 
with $\delta=2/3$ for ALICE Pb+Pb 2760~GeV data as upper (green) squares and PHENIX Au+Au 200~GeV data as 
lower (red) circles. The shaded band corresponds to the PHENIX data scaled by 
$(2760/200)^{0.06}$, including 
systematic uncertainties of $N_{\rm part}$.
}%
\label{ST}%
\end{figure}

Let us first observe that the equality of two estimates for $\alpha$, (\ref{first}) and (\ref{second}),
within an acceptable range of $n$, $\lambda$ and  $\lambda_{\rm ch}$,
cannot be obtained if $\lambda=\lambda_{\rm ch}$. As mentioned in Sec.~\ref{sec:Nch}
there is no theoretical understanding of this difference. One possibility might be the apparent
violation of GS that has been mentioned in Sec.~\ref{sec:basics}, namely the energy dependence 
of $S_T$. We do not have any experimental handle on this dependence, however 
recent Glauber model calculations \cite{Loizides:2017ack} show slight increase of $S_{T}$ with energy,
depicted in Fig.~\ref{ST}, which can be effectively parametrized as a power-law:%
\begin{equation}
S_{T}=\left(  \frac{W}{Q_{0}}\right)  ^{\lambda^{\prime}}\frac
{N_{\text{part}}^{\delta}}{Q_{0}^{2}}%
\label{STprime}
\end{equation}
with $\lambda'\sim 0.06$.
While (\ref{STprime}) may explain a difference in quality of GS for multiplicity vs. cross-section,
it is difficult to disentangle this dependence from the genuine energy dependence of $Q_s$ on $x$.
This is because we effectively include the energy dependence of $S_T$ in the process of tuning $\lambda$
in search for GS.  The naive inclusion of (\ref{STprime}) in Eqs.~(\ref{both}) and (\ref{Nch}) would only modify
the second equation for $\alpha$ in (\ref{second}) 
 shifting it to 1.29. However, a proper way of 
including the energy dependence of $S_{T}$ would require  a global fit with $\lambda$, $\lambda'$
and $\delta$ as free parameters, rather than adding $\lambda'$ in (\ref{second}).
%

There are many other sources that lead to the violation of GS in HIC, which we 
have ignored. For example, the factorization of the unintegrated gluon densities in (\ref{unintg}) 
into $A_{T}$ and $\phi$, neglecting the momentum dependence 
of $\alpha_{s}$
in (\ref{Nchdef}), as well as simplifying the assumptions about $N_{\rm part}$ dependence 
and the above-mentioned energy increase of $S_{T}$. It is obvious 
from Fig.~\ref{ST} that  these types of dependence do not factorize,
moreover, much better scaling is obtained with  the
power $\delta=3/4$. The latter, however, is of no  importance for our 
analysis, as $\delta$ drops out from the second equation of (\ref{conditions}). 
Finally, simple power-law dependence of the photon spectra is also a simplification that affects 
 the accuracy of our analysis. A choice  of more accurate data-driven functions could give an 
improved value of $\alpha$, closer to that extracted from experiment. This would require, however, a numerical 
analysis that we wish to avoid for the clarity of argumentation.
One should not forget that both scaling laws observed experimentally are only approximate.
This concerns also the charged particle spectra used here do calculate $dN_{\rm ch}/d\eta$ in (\ref{Nch}).

Even though one might argue that $\sim$ 10\% accuracy  is quite satisfactory, given a simplicity of the present analysis,
our results may indicate that there might be other components of direct photons that exhibit GS but have different
functional dependences than the initial stage component. This might resolve the difference between $\lambda$
and $\lambda_{\rm ch}$.
Identification of such components would
be of great interest.
 
 Our conclusion is that we have linked the multiplicity scaling and geometrical scaling
of direct photon $p_{T}$-spectra, estimating the scaling power $\alpha$ from simple assumptions 
based on the functional forms of the scaling functions $G$ and $F_{\gamma}$, as well as on the energy 
dependence and $N_{\rm part}$ dependence of the saturation scale. 
We have obtained two independent, albeit consistent, estimates of $\alpha$ that overshoot
experimental value by $\sim$ 10\%. This result relies on the fact that $\lambda \ne \lambda_{\rm ch}$.
We have argued that possible explanation of this difference, apart from trivial facts that charged
particles interact with the medium while photons do not and that integrated yields  of charged 
particles include non-scaling tails, might also indicate that other contributions of direct photon production 
exhibit GS but possibly have different functional forms than the one of the early stage production.


\bigskip

\section*{Acknowledgments}

We are grateful to Axel Drees and Larry McLerran for many important and fruitful discussions. 
The research of VK is supported under Department of 
Energy Contract No. DE-FG02-96-ER40988. 
The research of MP is supported by the NAWA (Polish National Agency for Academic Exchange) Bekker program and 
partially by the COST Action CA15213 THOR.
MP wishes to thank the Institute for Nuclear Theory at the University of Washington for its kind hospitality, 
stimulating research environment and for partial support by
 INT's U.S. Department of 
Energy Grant No. DE-FG02-00ER41132.

\end{document}